\documentstyle[aps,twocolumn,epsfig]{revtex}
\begin{document}

\draft

\title{Bending strain-driven modification of surface reconstructions: Au(111)}

\author{U.~Tartaglino$^{1,2,\ast}$,  E.~Tosatti$^{1,2,3}$,
        D.~Passerone$^{4,1,2}$, and F.~Ercolessi$^{5,1,2}$}
\address{
$^1$ SISSA, Via Beirut 2, I--34014 Trieste, Italy \\
$^2$ Unit\`a INFM/SISSA \\
$^3$ ICTP, Strada Costiera 11, I--34100 Trieste, Italy \\
$^4$ CSCS, Galleria 2, CH-6928, Manno, Switzerland \\
$^5$ Dip. di Fisica, Universit\`a di Udine, Via delle Scienze 208, I-33100 Udine, Italy
$^{\ast}$ FAX: +39-040-3787528; e-mail: tartagli@sissa.it
}
\date{\today}
\wideabs{
\maketitle
\begin{abstract}
Strain can affect the morphology of a crystal surface,
and cause modifications of its reconstruction 
even when weak, as in the case of mechanical bending. We carried out
calculations of strain-dependent surface free energy and
direct bending simulations demonstrating the change of incommensurate
reconstruction in Au(111) under strain, in good agreement with recent data.
Time-dependent strain should cause a sliding of the
topmost layer over the second, suggesting an interesting case
of nanofriction. Bending strain could also be used to 
fine tune the spacing of selectively absorbed nanoclusters.
\end{abstract}

%
}

Crystal surfaces often break their ideal bulk-like symmetry,
even in the absence of adsorbates, to lower their free energy
through the so-called reconstructions. Temperature, adsorbate coverage and
(in electrolytes) electrochemical potential \cite{ibach} cause
these reconstructions to evolve, either continuously
or through two-dimensional phase transitions\cite{persson}.
Work on semiconductor surfaces\cite{vanderbilt1,vanderbilt2,lagally}
indicated that external strain too can drive changes in surface
reconstruction. In this letter we
present theory and simulations indicating that reconstruction changes
at an incommensurate metal surface like Au(111) can be provoked by
application of strain, even with the weak strains attainable by bending
a platelet or slab. When compared with very recent data, which
appeared while this paper was being reviewed, the quantitative agreement 
is remarkable. We propose a possible application of this fine effect to 
selective adsorption of nanostructures on this surface.
Because the strain-induced reconstruction changes 
involve a continuous change of first-layer lateral atom density,
time-dependent bending strain will also drive a kind of incommensurate
nanofriction.

We introduce in a semi-infinite crystal a
uniform strain $\lambda\varepsilon_{\alpha\beta}$,  $\lambda$ increasing from
0 to 1. The surface free energy change, equalling the surface work, is
obtained through thermodynamic integration: 
\[
 \Delta ( \gamma A) = W_{s} =
 \int_{\mbox{A}} dA
 \int_0^1 \tau^{\mathrm{surf}}_{\alpha\beta}(\lambda)\,
                  \varepsilon_{\alpha\beta} d\lambda 
\]

where $\gamma$ is the surface free energy per
unit area, $A$ the area, $\tau^{\mathrm{surf}}_{\alpha\beta}$ the
(strain dependent) surface stress, and $\alpha$, $\beta$ span
the $(x,y)$ in-plane coordinates.
If the surface free energies (per unit area) of two different
reconstructions 1 and 2 satisfy $\gamma^{0}_{1} < \gamma^{0}_{2}$,
then in the absence of strain state 1 prevails over 2. However, their
respective surface stress tensors generally differ, and free energies
will thus vary with different slopes under strain, possibly
leading to $\gamma_1 > \gamma_2$. We are interested in a detectable 
modification of reconstruction periodicity, particularly if obtainable 
even with the weak strains provided by mechanical bending.

To explore that, we choose the promising example of Au(111)
as our test case. Au(111) displays a well
documented $(L \times \sqrt{3})$ reconstruction\cite{reconstr},
($L\approx 22.5$ in ordinary conditions). It is promising
because its long incommensurate period $L$ should readily
be changeable by a small but detectable amount. 
The $(L\times\sqrt{3})$ pattern,
commensurate along \mbox{$\langle 1\,1\,\overline{2}\rangle$},
is incommensurate in the orthogonal
\mbox{$\langle 1\,\overline{1}\,0\rangle$} direction, where the
reconstruction consists of a succession of
of hcp and fcc stripes, separated by sharp soliton--like boundaries, due to
an extra row of atoms for each period $L$ of reconstruction.
X-Ray diffraction experiments\cite{herringbone2}
show as temperature is raised from 300 to 800 K,
a \emph{continuous} shift of the periodicity $L$ from about 22.5 to 20.9,
which confirms the incommensurability. It also
indicates a denser and denser surface at higher temperatures,
a kind of negative expansion coefficient (probably arising,
as in the case of Au$(100)$ \cite{au100,au100bis}, to compensate
for a large outwards thermal relaxation of the first layer).
Here we shall describe how an external strain too can change the 
periodicity $L$ of Au(111).

To describe the initial $(L\times\sqrt{3})$ reconstruction we use
an empirical many-body classical potential, the glue
model\cite{glue}, known to describe reasonably well all the gold
surfaces. In the glue model, the ground state of Au(111) is properly
reconstructed, though with a somewhat shorter $T=0$ periodicity
$L=11$ \cite{bartolini}. The difference between that and the experimental
$L=22.5$ is a purely numerical discrepancy that can be corrected for, and of
no fundamental significance. The long range ``herringbone'' secondary
superstructure also well known on this surface \cite{herringbone2}
is much more delicate involving far smaller energies, and will not
concern us any further here. All calculations and simulations are conducted
for thin crystal slabs where strain is introduced by simulated
bending\cite{passero},
as in real-life experiments \cite{ibach} (although of course
our theoretical slabs are far thinner).

Calculations of the surface free energy change under strain start
with an $N$-layer fcc (111) slab, its two faces $(L \times \sqrt{3})$
reconstructed with different periodicities $L$ and $L'$. That is obtained 
by addition of one extra \mbox{$[1\,1\,\bar{2}]$} row of surface
atoms every $L$, followed by relaxation\cite{bartolini}.
We carry out molecular dynamics simulations of about 1\,ns at 100 K, to
obtain by standard methods\cite{RibarskyLandman} the average stress tensor
$\sigma_{\alpha\beta}$. From the same simulation we also extract 
a {\em layer--resolved} stress profile
$\tau_{\alpha\beta}(z_{l})$, defined as the force per unit length
acting in layer $l$:
\[
 \tau_{\alpha\beta}(z_l) =
  -\frac{1}{A}\left\langle \sum_{i\in \;l}
   \left(
    \frac{p_{\alpha i}p_{\beta i}}{m_{i}}
    + \frac{1}{2}\sum_{j} \frac{\partial U}{\partial r_{ij}}
    \frac{r_{ij,\alpha}r_{ij,\beta}}{r_{ij}}
   \right)
  \right\rangle \, ,
\]
where $i$ runs over the atoms of layer $l$, $j$ spans all the
atoms, and $U$ is the
potential energy, a function of the interatomic distances (this formula is
easily modified for application to a bent slab \cite{ugo}).
The elastic bending work is done in our case by $\tau_{yy}$,
$y\equiv [1\,\overline{1}\,0]$ being the strained direction,
orthogonal to the solitons.

\begin{figure}
 \epsfig{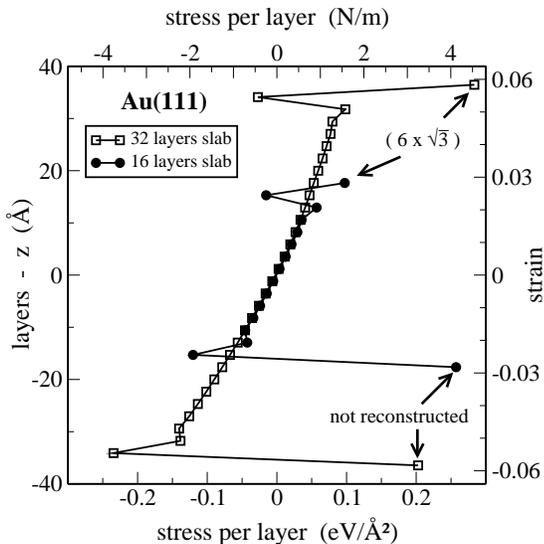}
 \caption{Layer--resolved stress profiles of two Au(111) slabs with 16 and 32
          layers, subjected to the same bending curvature. Note the large
          surface stress and the sub-surface oscillations. The non linear
          bulk stress profile is due to anharmonicity.}
\end{figure}
The bent slab stress profile obtained for $L=6$, $L'=\infty$ (at $T=100\,K$)
is shown in Fig.~1. The smoothly varying bulk-like stress in the interior
of the slab develops at the surface a sharp oscillation, swinging
from compressive in the second layer to tensile in the first layer.
As shown by comparison of two different slabs with the same curvature but
different thicknesses, $N=16$, and $32$, the surface stress of the thinner
slab can be immediately computed by subtracting off the corresponding
bulk-like stress of the thicker slab.
The bulk stress $\tau_{yy}^{\mathrm{(bulk)}}(z)$
deviates from the linearity for increasing strain in a way
that is very well described by a parabolic fit, the nonlinearity due to
anharmonicity. The layer--resolved surface stress is then obtained as
$ \tau_{yy}^{\mathrm{s}}(z_{l}) = \tau_{yy}(z_{l}) -
 \tau_{yy}^{\mathrm{(bulk)}}(z_{l})$.
Repeating calculations for increasing curvature $k$ we
obtain the curvature-dependent surface stress.
This approach, we note, will give free
energy differences, fully inclusive of the entropic term.
Finally the free energy change as a function of strain
is obtained by integrating \cite{ugo}:
\[
 W_{s}= \sum_{l\in \mathrm{layers}}
        \int_{0}^{k} d\kappa \; \tau_{yy}^{\mathrm{s}}(z_{l}) \,
        \frac{d\varepsilon_{yy}(z_{l})}{d\kappa}
\]
Likewise, the total surface stress can be obtained as
$\tau^{\mathrm{surf}}_{\alpha\beta}(k)= \sum_{l\in \mathrm{layers}}
\tau^{\mathrm{s}}_{\alpha\beta}(k,z_{l})$.

\begin{figure}
 \epsfig{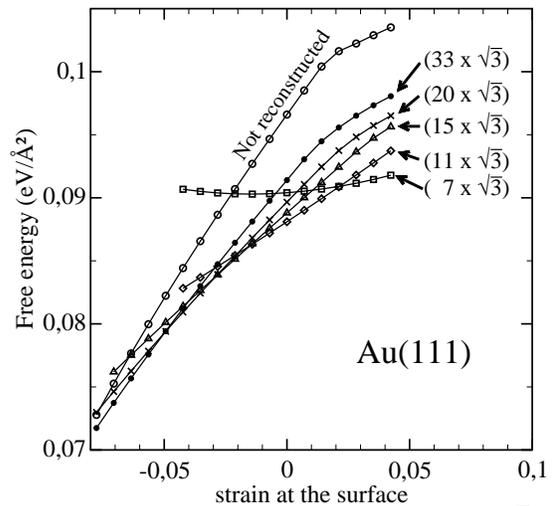}
 \caption{Free energies at 100 K of various $(L\times \protect\sqrt{3})$
          Au(111) surface reconstructions calculated as the strain
          is changed through bending. (The free energy values at zero
          strain are arbitrarily set at their independently
          obtained $T=0$ values). The predicted strain induced drift of $L$ can be extracted
          from the lowest envelope of these curves.}
\end{figure}

Fig.~2 displays our main result: the
strain--induced free energy variation of several different
$(L \times \sqrt{3})$ reconstructions of Au(111) at $T=100$ K.
The strain dependence found for free energy is relatively large, of
order 2 meV/\AA\ (about 1\% of the total surface free energy) for 1\% strain.
By comparison with the absolute zero-temperature surface energy
differences at zero strain, independently calculated (and used in
Fig.~2 to fix the unknown relative position of curves, thus neglecting 
an entropy term, hopefully small at low temperatures)
we find that 0.18 meV/A is roughly the surface free energy difference
between periodicities $L$ and $L+1$ for $L$ around 20 in Au(111).
By interpolating Fig.~2 we find that a change of periodicity from, 
say, $L=23$ to $L=24$ can in principle be provoked by a compressive
surface strain of order 0.2\%.
More generally, strain will cause a periodicity drift determined by
the lower envelope of all the curves: positive (tensile) strain
decreasing $L$, negative (compressive) strain increasing it.
Eventually for a theoretically very large strain $L=\infty$ would be 
reached, where the reconstruction itself disappears. That strain, of 
order $-2\%$ in Au(111) (see below), might
however be too large to be achieved by bending.

Our predictions, based on free energies, find a first confirmation
in a computer experiment, consisting of a realistic molecular dynamics
simulation, based on introducing a bending strain in the same slab 
geometry previously used for the surface stress calculation. If we
start out with two identically reconstructed surfaces, bending strain should
cause $L$ to spontaneously increase on the concave face, and decrease on the convex one.
The simulation must however overcome a technical problem, connected
with particle conservation. A change of reconstruction periodicity $L$
implies a change of first--layer lateral surface density,
equal to $(1+1/L)$ times the density of an unreconstructed layer.
The initially flat, defect free reconstructed surface should, under
bending, spontaneously expel atoms from the top layer to form islands on
the concave face, while opening craters that expose a portion of the layer
below on the convex face\cite{passerocrater}. Such phenomena are slow,
needing times much longer than the typical simulation nanosecond.
Because of that, it would seem that molecular dynamics simulation should
be abandoned. A practical way out of this problem
\cite{twostep} is to include a loose grid of steps
in the starting configuration, which amounts to simulate a
very close vicinal face rather than the original
\mbox{(1\,1\,1)} surface.
\begin{figure}
 \epsfig{file=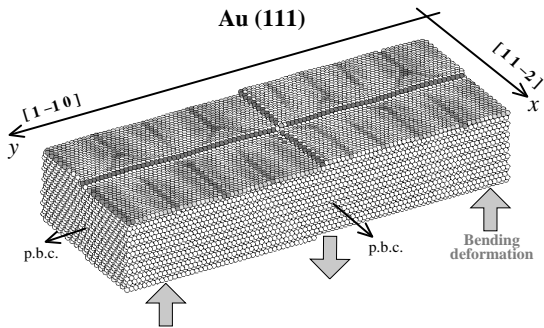, width=0.4\textwidth}
 \caption{The upper half of the simulation slab, with two orthogonal steps.
          Periodic boundary conditions require no extra steps at the
          boundaries. The grayscale reflects atomic coordination: atoms at
          the solitons between fcc and bcc regions appear darker.}
\end{figure}
Our realization, shown in Fig.~3, comprises
four sub-terraces of heights 0,\,1,\,1,\,2, separated by steps, and
connected one with another across suitably shifted boundary
conditions\cite{twostep}. No other steps occur at the boundaries,
hence the four sub-terraces constitute in reality a single large terrace.
The terrace size used is $88d \times 20\sqrt{3}d$, with $d$ the nearest
neighbour distance. With this geometry we can carry out
various simulations, bending an initially flat surface and unbending
an initially bent surface. In order to speed up kinetics, the temperature
is raised well above room temperature, up to about $T=600 K$.

\begin{figure}
 \epsfig{file=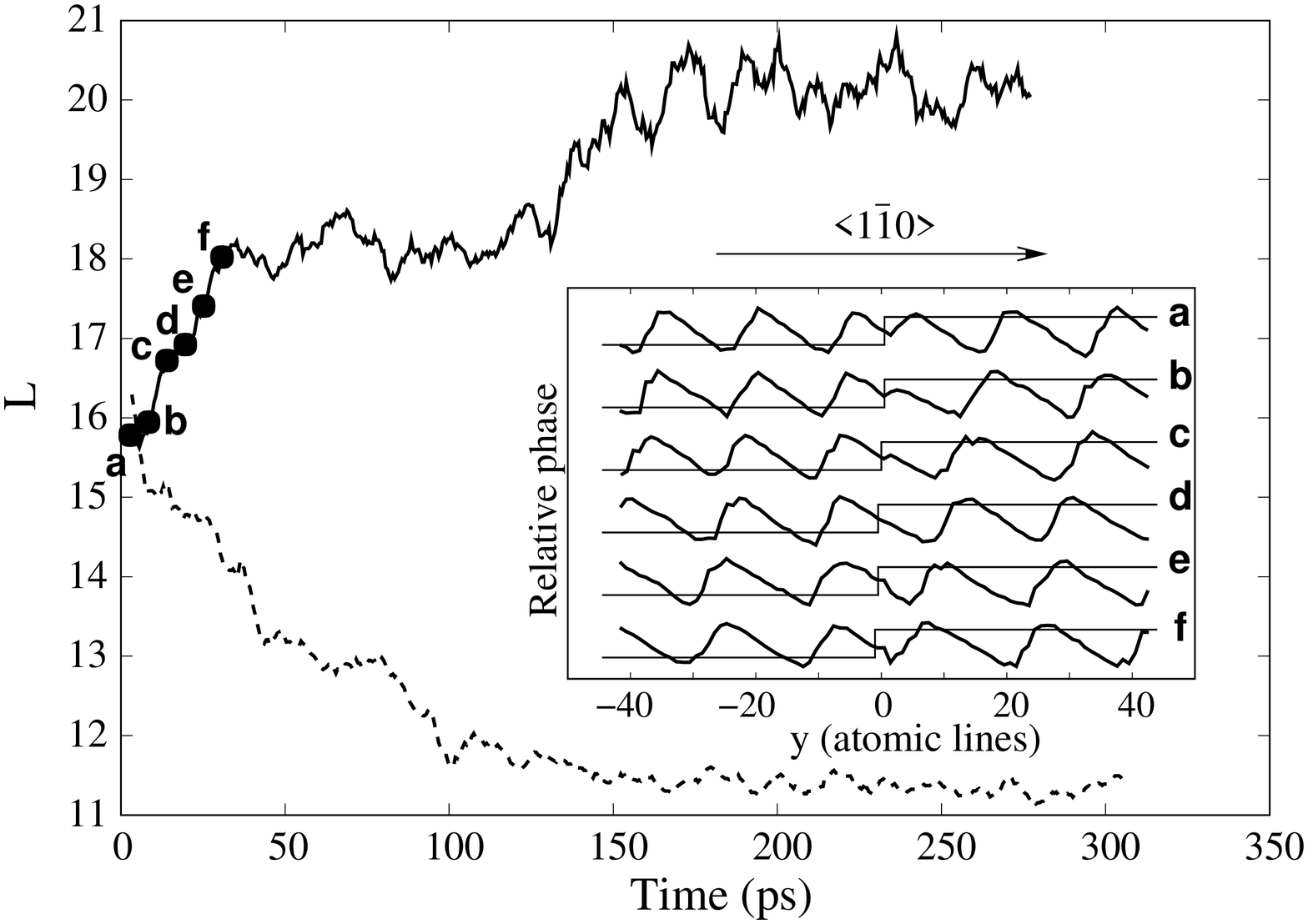, width=0.4\textwidth}
 \caption{Time evolution of the reconstruction periodicity, from a
          nonequilibrium starting point with $L=16$, towards equilibrium.
          Solid line: bent slab with a compressive surface strain of
          $-4.7 \%$. Dashed line: flat slab with zero surface strain.
          Geometry as in Fig. 3, T= 800 K. Inset: detail of the
          phase difference between the atoms of the first two layers, at
          selected instants of time during the bending. Note that the step,
          indicated schematically at each of the instants, has moved
          slightly to the left in correspondence with the increase of $L$
          (decrease of lateral density). The excess soliton is absorbed
           at the step upper rim between instants (a) and (c).}
\end{figure}
Representative simulation results are illustrated in Fig.~4, where the
continuous drift of reconstruction periodicity for a surface initially
reconstructed with $L$  around 16 is demonstrated for two different situations,
a large compressive surface strain, and zero strain respectively, the former
obtained through simulated bending. The periodicity drift from 16 to
about 20 and from 16 to about 11 in the two cases is in good agreement
with the free energy prediction of Fig.~2 .
The inset details the behavior of the phase of the topmost atom
layer lattice relative to the bulk-like second layer (modulo $2\pi$)
as seen on the bent slab at selected instants of time.
Data are averaged along the \mbox{$[1\,1\,\bar{2}]$} direction
to reduce the noise due to thermal excitation: averaging explains
why the phase jumps around $2\pi$ are replaced by a smooth
behaviour. We find that the periodicity changes are actuated through the
appearance (or disappearance) of solitons, nucleated
preferentially at the upper rim of the surface step, where
atoms possess the lowest coordination.

Experimentally, attempts to provoke and observe by STM  surface
reconstruction modifications by bending, have been carried out
recently, precisely on Au(111). Because the strains attained are small
they were found to affect mostly the secondary structure.
An early unpublished report by Huang et al.\cite{huang},
described strain--induced removal of the wrongly oriented domains
of the secondary herringbone structure
with strains as small as $1.3\times 10^{-5}$, with occasional removal
of the whole reconstruction. Newer data of Schaff et al.\cite{schaff}
show that the secondary structure is again removed but the $(L\times\sqrt{3})$
reconstruction survives at least up to a maximum compressive strain of
$0.4\%$. By closer examination of their data we note that under strain,
exerted at an angle of $30^{\circ}$ with respect
to the solitons, there is evidence of a drift of periodicity $L$ from
about $23$ to about $25$.
We can directly compare that with our calculations. We find that the
calculated $L$-dependent surface free energy for variable bending is well 
fitted by a Pokrovskii-Talapov form
\cite{persson} $\Delta E = -a/L+b/L^3$, where $a$ is the cost of 
solitons and $b$ reflects their mutual interaction $\propto 1/L^2$.
Here $a(\varepsilon)=a'\cdot (\varepsilon-\varepsilon_{c})$
will vanish at a critical strain $\varepsilon_{c}$, so that
$L=K(\varepsilon-\varepsilon_{c})^{-1/2}$, where $K=\sqrt{3b/a'}$.
Based on our computed surface energies and anisotropic surface s
tress tensor, we derived $K$ and $\varepsilon_{c}$
for arbitrary strain direction, so as to compare the experimental
drift with calculated periodicity changes (Fig.~5).
The agreement obtained is very encouraging,
fully supporting our qualitative expectations, and demonstrating the
applicability of the proposed mechanism of surface modification
under bending strain.
\begin{figure}
 \epsfig{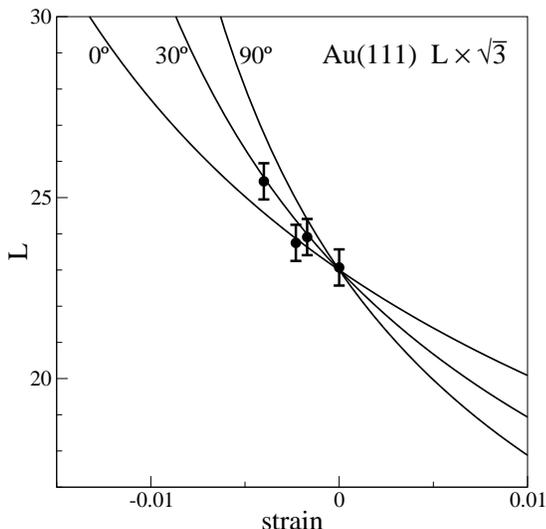}
 \caption{Change of Au(111) reconstruction period $L$ as a function of applied
          strain. 
          Solid lines, are theoretical predictions, in the form
          $L=K(\varepsilon-\varepsilon_d)^{-1/2}$, for strains
          orthogonal ($\theta=90^{\circ}$, $K=2.84$,
          $\varepsilon_d=-0.0153$), oblique
          ($\theta=30^{\circ}$, $K=3.34$, $\varepsilon_d=-0.0210$)
          and parallel ($\theta=0^{\circ}$, $K=4.12$,
          $\varepsilon_d=-0.0323$) to the \mbox{$[1\,1\,\bar{2}]$} oriented 
          solitons . These parameters best fit
          our simulation results, with a shift in strain, to 
          make the $L=23$ point correspond to strain $\varepsilon=0$.
          Data points were extracted from the STM images of
          ref.\ \protect\cite{schaff}, where the strain direction was
          $\theta=30^{\circ}$.
         }
\end{figure}

The strain-induced reconstruction changes described above can 
lead to potentially interesting consequences and applications. 
In a surface that is perturbed by a time-dependent strain, there will be
as a consequence a sliding nanofriction of the incommensurate
reconstructed top layer of Au(111) against the regular
second layer. For example, an oscillatory bending would
result according to Fig.~4 in a corresponding variation of $L$, driving
an oscillatory sliding of the first layer onto the second.
A pinned state of the top layer will imply
a threshold strain magnitude to actuate the sliding. 
When the sliding occurs, a surface nanofrictional contribution
should appear as a part of the mechanical damping of a Au(111) platelet.
Among the applications, there could be the possibility to tune
through bending strain the spacing between deposited nanoclusters.
Recently, deposition of 1-nitronaphtalene (NN)
nanoclusters on a stepped Au(111) surface 
has revealed 
\cite{tonidevita}
that at the low coverage adsorption at the step
is favored in the fcc zones of the $L \times \sqrt 3$ reconstruction pattern
and unfavored in the hcp zones, and in this way, stripes can be produced.
The distance among the stripes could in principle be changed upon bending, with
interesting consequences in the field of miniature device fabrication.

This work was supported through contracts INFM PRA NANORUB,
and MIUR COFIN. We thank P. Zeppenfeld and O. Schaff for
bringing their work to our attention, and for illuminating discussions.


\begin{thebibliography}{99}

\bibitem{ibach} H.Ibach, Surf. Sci. Reports \textbf{29}, 193 (1997)
\bibitem{persson} B.N.J.Persson, Surf. Sci. Reports \textbf{15}, 1 (1992)
\bibitem{vanderbilt1} D.Vanderbilt, Phys. Rev. Lett. \textbf{59}, 1456 (1987)
\bibitem{vanderbilt2} D.Vanderbilt, Phys. Rev. \textbf{B R36}, 6209 (1987)
\bibitem{lagally} F.Liu and M.G.Lagally, Phys. Rev. Lett.
 \textbf{76}, 3156 (1996)
\bibitem{reconstr}
 Y.Tanishiro, H.Kanamori, K.Takayanagi, K.Yagi, and G.Honio, Surf. Sci.
 \textbf{111}, 395 (1981); U.Harten, A.M.Lahee, J.P.Toennies, and Ch.W\"{o}ll,
 Phys. Rev. Lett. \textbf{54}, 2619 (1985), and references therein.
\bibitem{herringbone2} A.R.Sandy, S.G.J.Mochrie, D.M.Zehner, K.G.Huang
 and D.Gibbs, Phys. Rev. B \textbf{43}, 4667 (1991)
\bibitem{au100} K.Yamazaki, K.Takayanagi, Y.Tanishiro, and K.Yagi,
 Surf. Sci. \textbf{199}, 595 (1988)
\bibitem{au100bis} D.Gibbs, B.M.Ocko, D.M.Zehner and S.G.J.Mochrie,
 Phys. Rev. B \textbf{42}, 7330 (1990)
\bibitem{glue}  F.Ercolessi, M.Parrinello, and E.Tosatti,
 Phil. Mag. A\textbf{58}, 213 (1988)
\bibitem{bartolini} F.Ercolessi, A.Bartolini,  M.Garofalo, M.Parrinello,
 and E.Tosatti, Surf. Sci. \textbf{189/190}, 636 (1987).
\bibitem{passero} D.Passerone, E.Tosatti, G.L.Chiarotti, and F.Ercolessi,
 Phys. Rev. B \textbf{59}, 7687 (1999)
\bibitem{RibarskyLandman} M. W. Ribarsky and U. Landman,
 Phys. Rev. B\textbf{38}, 9522 (1988)
\bibitem{ugo} U. Tartaglino, D. Passerone, E. Tosatti and
 F. Di Tolla, Surface Science, \textbf{482-485}, 1331 (2001)
\bibitem{passerocrater} D.Passerone, F.Ercolessi, and E.Tosatti, Surf. Sci.
 \textbf{377--379}, 27 (1997)
\bibitem{twostep} D.Passerone, F.Ercolessi, U.Tartaglino and E.Tosatti,
 Surf. Sci. \textbf{482-485}, 418 (2001)
\bibitem{huang} L.Huang, P.Zeppenfeld, and G.Comsa, private communication
\bibitem{schaff} O.Schaff, A.K.Schmid, N.Bartelt, J.de la Figuera and
 R.Q.Hwang, Mat. Sci. Eng. \textbf{A 319-321}, 914 (2001)
\bibitem{tonidevita} M.Vladimirova, M.Stengel, A.De Vita, A.Baldereschi,
 M.B\"ohringer, K.Morgenstern, R.Berndt, and W.-D.Schneider,
 Europhys. Lett. \textbf{56}, 254 (2001)

\end{thebibliography}
\end{document}